\documentclass[amsmath,amssymb,aps,prb,floatfix,]{revtex4-1}

\usepackage[dvips]{graphicx}
\usepackage{youngtab}
 
\begin{document}
 
\title{BCS models of Josephson qubits I. Energy spectra}
 
\author{Robert Alicki}
\email{fizra@univ.gda.pl}
\author{Wies\l aw Miklaszewski}
\email{fizwm@univ.gda.pl}
\affiliation{
Institute of Theoretical Physics and Astrophysics, University
of Gda\'nsk,  Wita Stwosza 57, PL 80-952 Gda\'nsk, Poland}
 
\date{\today}
 
\begin{abstract}
There exists a large number of experimental and theoretical results supporting the picture of \emph{macroscopic qubits} implemented by nanoscopic Josephson junctions of three different types -- \emph{charge qubit}, \emph{flux qubit} and \emph{phase qubit}. The standard unified description of such systems is based on the formal quantization of the phenomenological Kirchhoff equations for the corresponding circuits. In this paper  a simplified version of the BCS  theory for superconductors is used to derive microscopic models for all types of small Josephson junctions. For these models the state-dependent individual tunneling of Cooper pairs couples \emph{ground pair} states with \emph{excited pair} states  what leads to a more complicated  structure of the lowest lying energy levels. In particular, the  highly degenerate levels emerge, which act as probability sinks for the qubit. These models allow also for the coupling to phonons as an efficient mechanism of relaxation for all types of junctions. The alternative formulas concerning basic spectral parameters of superconducting qubits are presented and compared with the experimental data. Finally, the question whether small Josephson junctions can be treated as  \emph{macroscopic quantum systems} is briefly discussed.
\end{abstract}
 
\pacs{}
\maketitle
 
\section{Introduction}

In the last decade remarkable experiments were performed involving measurements and manipulations of states for  a single or several nanoscopic Josephson junctions (JJ) which were consistently interpreted in terms of two level quantum systems\cite{W,Cla,Nak,Leh,Gui,Bla,Mar,Stef,Hof,Ans,Wal,Fri}. The main assumption in the theoretical analysis is that such a many-body mesoscopic system can be effectively treated as a quantum system of a single degree of freedom typically described by a large spin or nonlinear oscillator model. The standard construction of the quantum Hamiltonian for different types of JJs is based on the formal quantization of the Kirchhoff equation for the corresponding macroscopic circuit\cite{W}.   Another approach uses an effective picture of Bose-Einstein condensate which represents Cooper pairs at low temperatures\cite{Jak}. More fundamental derivations based on path-integral techniques lead to essentially equivalent models as well\cite{Eck, Kor}. The simple representative of this class of single degree of freedom models is the following Josephson Hamiltonian  describing the effects of Coulomb repulsion and tunneling for a Cooper pair box (CPB)\cite{Rod}
\begin{equation}
\hat{H} = 4E_C\left(\hat{J}_3- n_g\right)^2  - \frac{E_J}{2j}\hat{J}_1.
\label{cpbham1}
\end{equation}
Here, $\hat{J}_k$, $k=1,2,3$, are spin operators for the spin-$j$ such that the number of Cooper pairs at  equilibrium is close to $j$, $E_C$ is the charging energy, $E_J$ is the Josephson energy characterizing the tunneling magnitude, and $n_g$ is a control parameter. 

The obtained nonlinear Hamiltonians yield the structure of two lowest energy levels which at the low enough temperatures can be separated from the others to form an effective \emph{macroscopic qubit}. In particular for the Hamiltonian (\ref{cpbham1}) and under the condition $E_C\gg E_J$ these states are approximatively spanned by the eigenstates  $|m_0\rangle$ and $|m_0+1\rangle$ of the operator $\hat{J}_3$ with $ m_0 \le n_g \le m_0+1$.

The main problem with such models is the presence of a typically strong and collective
coupling to an environment. Namely, it is  expected that the observed states should be  rather well-localized semiclassical ones (like coherent states for the model (\ref{cpbham1})), which seem to be the only relatively stable with respect to  external noise\cite{Ben}. 
However, the semiclassical states for a model of small JJs  are characterized  by large charge fluctuations which are not observed in the experiments with CPBs . Therefore, either environmental decoherence producing semiclassical states does not work for Josephson qubits at the typical time scale of the experiments or the standard single degree of freedom model is not correct. The first alternative seems to be unlikely because the semiclassical character of observed states is confirmed in the recent experiments on atomic Bose-Einstein condensate (BEC) in a double-well potential\cite{Es}. Although this is a physically different system its mathematical description is the same as for the standard model of a CPB and given by the Hamiltonian  of the form (\ref{cpbham1}). Therefore, we follow the second possibility and propose an essentially modified theoretical description of small JJs. 
 
In Sec. II two basic approximations for the Bardeen-Cooper-Schrieffer (BCS) Hamiltonian: Bogoliubov-Valatin model Hamiltonian and the collective spin models are discussed. It is argued, that with the specific choice of model parameters different from the standard ones but consistent with phenomenology of superconductivity, the collective spin model better describes physics of small superconducting devices. In particular, this model predicts the existence of specific excitations called \emph{excited pairs} which can be produced by the individual  tunneling or scattering from the \emph{ground pair} states. One should stress, that usually excited pairs are not considered in the literature on JJs, in contrast to the single electron/hole quasi-particles. Moreover, in the standard derivation the collective character of scattering/tunneling is also assumed.

In Sec. III this new picture is applied to CPB, in Sec. IV to \emph{flux qubit} (FQ ), and in Sec. V to current biased junction (CBJ). The main feature is the appearance of the additional energy levels, among them highly degenerate ones. This level structure essentially modifies the \emph{qubit picture} of small JJs and  the dynamics of their relaxation.  Section VI is devoted to the comparison of predictions of the presented unifying microscopic model with experimental data for all types of superconducting qubits. The advantage of the presented theory is that the approximative formulas for the qubit frequencies contain less free parameters than the standard ones but nevertheless agree with the data.

\section{BCS model and its approximations}

The simplest version of the Hamiltonian that incorporates pairing interaction and reproduces not only the basic phenomenology of superconductivity\cite{Thou} but is successfully applicable to small superconducting grains\cite{Del} is the following
\begin{equation}
\hat{H} =\hat{H}_0 + \hat{H}_{\mathrm{red}},
\label{BCS}
\end{equation}
where
\begin{equation}
\hat{H}_0 = \sum_{k,\sigma=\pm} (\epsilon_k -\epsilon_F)\, \hat{c}^{\dagger}_{k\sigma}\hat{c}_{k\sigma}
\label{H0}
\end{equation}
and
\begin{equation}
\hat{H}_{\mathrm{red}} = - \frac{g}{K} \sum_{k,l}  \hat{c}^{\dagger}_{k+}\hat{c}^{\dagger}_{k-}\hat{c}_{l-}\hat{c}_{l+}.
\label{Hred}
\end{equation}
Here we denote by $|k,\pm\rangle$ a single electron basis of pairs of time-reversed states enumerated
by a discrete index $k= 1,2,\ldots,K$ and by $\hat{c}_{k\pm}$, $\hat{c}^{\dagger}_{k\pm}$ the associated set of fermionic annihilation and creation operators. Those single electron states correspond to eigenstates with the single electron energy levels $\epsilon_k $ within a cut-off $\hbar\omega_{cut}$ around the Fermi energy $\epsilon_F$. The coupling constant
$g > 0$ has here a dimension of energy\cite{g}.

Although there exists a remarkable exact solution for this Hamiltonian\cite{Ric} it is too complicated for our purposes. In the following we briefly discuss and compare two well known approximative schemes. 

\subsection{Bogoliubov-Valatin model Hamiltonian}

The basic idea of this approach is to replace the  Hamiltonian (\ref{BCS}) containing four-body interaction (\ref{Hred}) by a two-body (mean-field) one
\begin{eqnarray}
\hat{H}_{mf}&=& \sum_{k,\sigma=\pm} (\epsilon_k -\epsilon_F)\, \hat{c}^{\dagger}_{k\sigma}\hat{c}_{k\sigma}\nonumber\\
&&- \sum_{k} \left(\Delta \hat{c}^{\dagger}_{k+}\hat{c}^{\dagger}_{k-}+\Delta^\ast \hat{c}_{k-}\hat{c}_{k+}\right)
\label{Hmf}
\end{eqnarray}
with the \emph{gap parameter} $\Delta $ determined using a  self-consistent 
averaging
\begin{equation}
\Delta = \frac{g}{K}\sum_k\langle \hat{c}_{k-}\hat{c}_{k+}\rangle_{\mathrm{av}}.
\label{gap}
\end{equation}
In particular one can put the model Hamiltonian (\ref{Hmf}) into the expression for the thermal average  to obtain the value of $\Delta $ as the solution of the \emph{gap equation}. It can be done because the Hamiltonian (\ref{Hmf}) can be transformed to the form of the Bogoliubov-Valatin free Hamiltonian
for the \emph{quasi-particles} 
\begin{eqnarray}
\hat{H}_{BV}&=& \sum_{k,\sigma=\pm} E_k \hat{C}^{\dagger}_{k\sigma}\hat{C}_{k\sigma},
\label{Hfree}
\end{eqnarray}
where $E_k= \sqrt{(\epsilon_k - \epsilon_F)^2 + \Delta^2}$, described by a set of new  fermionic annihilation and creation operators $\hat{C}_{k\pm} ,\hat{C}^{\dagger}_{k\pm}$ being linear combinations of the old ones. The gap parameter $\Delta$ can be chosen positive and the gap equation for the thermal average in the grand canonical ensemble determines the temperature dependence  $\Delta(T)$. One should notice that the gap equation makes sense only for the reference states with indefinite number of Cooper pairs, like grand canonical ensemble or coherent-type states which display essentially normal fluctuations.

The fundamental results of this theory are the relations between the critical temperature
$T_c$ ($\Delta(T_c) = 0$)  and the parameters of the model. In terms of the parametrization used
here they read\cite{Tin}
\begin{eqnarray}
\Delta(0)&=& 1.76k_B T_c,\\
\label{gap0}
k_B T_c &=& 1.13\, \frac{\hbar\omega_{cut}}{2\sinh{\hbar\omega_{cut}/g}}.
\label{gap1}
\end{eqnarray}
Only the first relation has been tested in many experiments and found to be reasonable while the second one strongly depends on the unknown parameter $\omega_{cut}$.

As shown already by Thirring and Wehrl\cite{Thi1,Thi2} the predictions  based on the model Hamiltonian (\ref{Hfree}) become exact only in the thermodynamic limit for the grand canonical ensemble when the fluctuation terms of the order $\sqrt{K}$ become negligible in comparison to the bulk terms proportional to $K$. Therefore, one cannot expect that this model gives precise structure of few lowest lying energy levels in the case of mesoscopic samples with essentially fixed number of Cooper pairs.

\subsection{The choice of parameters}
 
In the modern literature one usually identifies the cut-off frequency $\omega_{cut}$ with the Debye frequency $\omega_D$\cite{Tin} ($\hbar \omega_D \gg k_BT_c$) what due to (\ref{gap1}) implies 
\begin{equation}
\hbar \omega_D = \hbar \omega_{cut}\gg g \gg k_BT_c.
\label{gap2}
\end{equation}
In the original BCS paper\cite{BCS} the choice  $\hbar\omega_{cut}\simeq k_B T_c$  was suggested what implied that all relevant energy parameters of the model could be of the same order of magnitude
\begin{equation}
\Delta(0)\simeq k_B T_c \simeq \hbar\omega_{cut} \simeq g.
\label{gap3}
\end{equation}
The choice of $\omega_{cut}$ determines important, although not directly measurable, quantities. The first one is the actual magnitude of the coupling constant $g$. The second one is the number of Cooper pairs in the sample at zero temperature, given by the formula $K/2= \hbar\omega_{cut} N(0)$ where $N(0)$ is the density of electronic states at the Fermi surface (spin is not counted). Hence the first choice (\ref{gap1}) yields this number larger by two orders of magnitude than the second one (\ref{gap2}). 

The are few heuristic arguments in favor of the choice (\ref{gap3}):
\begin{enumerate}
\item It implies that there is a single energy scale determining the superconducting phenomena given by the magnitude of the electron-phonon coupling, while the choice (\ref{gap2}) involves two quite different energy scales ($\hbar \omega_D \gg k_BT_c$). In a general case of particle interaction mediated by bosons the cut-off in boson momentum $p_{cut}$ influences the magnitude of interaction for interparticle distances  $d\leq \hbar/p_{cut}$. In the case of phonons this yields the distances $d\leq a$ ($a$ is the lattice constant). On the other hand, the "size" of a Cooper pair is at least two orders of magnitude larger than $a$ and hence the Debye cut-off should not enter the effective interaction between Cooper pairs
in the BCS Hamiltonian (\ref{Hred}).
\item For a normal state of a metal the approximative number of thermally excited electrons is $\sim k_B T N(0)$ while the others are "frozen in a Dirac sea".  When the temperature decreases to the critical value  $T_c$, it is plausible to expect that only the  $\sim k_B T_c N(0)$ excited electrons begin to form Cooper pairs and the others remain "frozen". Finally, when the temperature approaches zero all of them recombine into Cooper pairs. Hence, the number of Cooper pairs at zero temperature given by $K/2$ should be rather of the order of $k_B T_c N(0)$ than $\hbar \omega_D N(0)$.
\end{enumerate}

\subsection{Collective spin model of superconductor}

In order to produce a simple, exactly solvable model Anderson\cite{And} and independently Wada \emph{et al.}\cite{Wada} considered a simplification of the Hamiltonian (\ref{BCS}) neglecting the kinetic energy term $\hat{H}_0$.

Introducing the pair operators
$\hat{b}_k =\hat{c}_{k-}\hat{c}_{k+}$ and $\hat{b}^{\dagger}_k ={c}^{\dagger}_{k+}\hat{c}^{\dagger}_{k-}$ one can treat the system as \emph{hard-core bosons}
or equivalently as a system of $K$ spins-$1/2$ with spin operators $\hat{s}^{\alpha}_k$, $\alpha = x,y,z$, such that
\begin{eqnarray}
\hat{b}_k& =&\hat{s}^x_k -i\hat{s}^y_k=\hat{s}^-_k, \nonumber\\
\hat{b}^{\dagger}_k&=&\hat{s}^x_k +i\hat{s}^y_k=\hat{s}^+_k, \label{spin1}\\
\hat{b}^{\dagger}_k\hat{b}_k&=& \hat{s}^z_k +\frac{1}{2}.\nonumber
\end{eqnarray}
Defining the collective spin  operators $\hat{\mathbf{J}}= (\hat{J}_x ,\hat{J}_y,\hat{J}_z )$ and $\hat{J}_{\alpha}= \sum_k\hat{s}^{\alpha}_k$ one can write the Hamiltonians
\begin{eqnarray}
\hat{H}_0& =& \sum_{k} 2(\epsilon_k -\epsilon_F)\hat{b}^{\dagger}_{k}\hat{b}_{k}=\sum_{k} 2(\epsilon_k -\epsilon_F)\hat{s}^z_{k} + \mathrm{const},\nonumber\\
\hat{H}_{\mathrm{red}}& =& - \frac{g}{K} \left(\hat{\mathbf{J}}^2 - \hat{J}_z^2 + \hat{J}_z\right)
\label{bcs1}
\end{eqnarray}
which are equivalent to the Hamiltonians (\ref{BCS}-\ref{Hred}) only on the states  not including excitations in the form of unpaired electrons. In the following we assume that $\epsilon_k \simeq \epsilon_F$ and use  $\hat{H}_{\mathrm{red}}$  (\ref{bcs1}), called \emph{strong coupling limit Hamiltonian}, as the approximative Hamiltonian of the system.
 
The $K$-spins Hilbert space can be decomposed into subspaces corresponding to irreducible representations of $SU(2)$ of the dimension $2j+1$ and multiplicity $r_j$ represented by the suitable Young frames\cite{Bar}
\begin{equation}
\mathbb{C}^{2^K} = \bigoplus_{j=0(1/2)}^{K/2} \mathbb{C}^{2j+1}\otimes\mathbb{C}^{r_j} .
\label{hilbert}
\end{equation}
One can use as eigenvectors of $\hat{H}_{\mathrm{red}}$ the orthonormal basis $|j,m;r\rangle$
\begin{eqnarray}
\mathbf{J}^2|j,m;r\rangle &=& j(j+1)|j,m;r\rangle,\nonumber\\
\hat{J}_z|j,m;r\rangle & = & m|j,m;r\rangle,
\label{basis}
\end{eqnarray}
where $r=1,2,\ldots,r_j$, to obtain the corresponding eigenvalues of $\hat{H}_{\mathrm{red}}$
\begin{equation}
E_{jm} = -\frac{g}{K}\left( j(j+1)- m(m-1)\right).
\label{evalues}
\end{equation}
For a fixed total number of Cooper pairs $N=K/2 + m$ and hence fixed $m$ the single ground state is given by a nondegenerate eigenvector $ |K/2,m\rangle$. The highly degenerate states with $j= K/2 - p$, $p= 1,2,\ldots$, describe $p$ excitations called
\emph{excited pairs } in the original BCS paper\cite{BCS}. They are still composed of  Cooper pairs  but their wave functions possess different symmetry with respect to  permutations of pairs (ground state is completely symmetric) given by the corresponding Young tables.
This simplified model  with an additional structure which takes into account single electron excitations has been studied at
finite temperatures\cite{Thou}. 
\subsection{Validity of the collective spin model}

In the following we argue that for the mesoscopic samples the structure of the lowest lying levels is reasonably well-described by the Hamiltonian (\ref{bcs1}) with the kinetic part  $\hat{H}_0$ treated as a "small" perturbation of $\hat{H}_{\mathrm{red}}$. This is consistent with the choice (\ref{gap3}) and implies the relatively low number of Cooper pairs $K/2 \simeq N(0)k_B T_c$.

In the case of a small electrode, when the Coulomb repulsion should be included, the dominating part of the Hamiltonian reads
\begin{equation}
\hat{H}^C_{\mathrm{red}} =  - \frac{g}{K} \left(\hat{\mathbf{J}}^2 - \hat{J}_z^2 + \hat{J}_z\right)+ 4E_C \left(\hat{J}_z- \bar{m} \right)^2 .
\label{hamcpb1}
\end{equation}
Here $E_C = 2e^2/C$ is the charging energy related to the capacitance $C$, and the parameter $\bar{m}$, $|\bar{m}|\ll K$, determines the average excess number of Cooper pairs in the system. In the case of relatively large system, i.e., for $E_C \ll g$, the lowest lying states are of the form $ |K/2,m\rangle$ with $|m - \bar{m}|\ll K$. Therefore, the relevant Hilbert space can be represented by a \emph{highest spin} Hilbert space of the dimension $K+1$ and the Hamiltonian has the same form as (\ref{hamcpb1}) with the collective operators restricted to this subspace. The highest spin Hilbert space is invariant under the action of collective operators $\{\hat{J}_{\alpha}, \alpha = x,y,z\}$. The same  holds if the external action on the system is described by the Hamiltonian  being a function of $\{\hat{J}_{\alpha}\}$ what leads to a  \emph{large spin model} of a mesoscopic JJ. This model is essentially equivalent to all  single-degree of freedom models of CPB used in the literature\cite{Rod}. However, for $E_C \ge g$ or/and non-collective interactions with an environment
the states with $j < K/2$ containing excited pairs become important. 

For the illustration consider the ground state and the lowest excited levels of the Hamiltonian (\ref{hamcpb1}) with
$E_C \gg g$, the even value of $K$ and $\bar{m} = 0$. The ground state has form $|K/2,0\rangle$ with the energy
$-g(K/4 + 1/2)$ and the first excited state is $(K-1)$-fold degenerate $|K/2-1,0; r\rangle$, $r= 2,\ldots,K$, with the energy $-g(K/4 - 1/2)$ and hence separated from the ground one by the energy gap $g$. The Young tables corresponding to the ground state (upper one) and to the first excited states labeled by $b= 2,3,...,K$ (lower one) are as follows
\begin{eqnarray}
&\young(1\cdot\cdot abc\cdot\cdot K)\nonumber\\
&\young(1\cdot\cdot ac\cdot\cdot K,b)
\end{eqnarray}
To describe the excitations in the form of unpaired electrons one has to build a more complicated Hilbert space being a direct sum of the Hilbert spaces $\mathbb{C}^{2^{K'}}$. Here $K'= K -2p$ corresponds to the different sets of electronic states $|k,\pm\rangle$ with $p$ pairs excluded by the \emph{Pauli blocking effect}\cite{Del}. This extended model has been analyzed at finite temperatures\cite{Thou} and the main predictions  can be summarized as
\begin{equation}
\Delta(0)= g = 2k_B T_c.
\label{gap4}
\end{equation}
In the following we assume also that
\begin{equation}
\hbar\omega_{cut} = g =2 k_B T_c
\label{gap5}
\end{equation}
what is consistent with the BCS prediction (\ref{gap1})

The numerical values of the important measurable relations obtained from the collective spin model differ slightly from the BCS values. These deviations are due to neglecting  the kinetic energy term which, even in the
regime (\ref{gap3}), is still comparable to the pairing energy. Nevertheless, the collective spin model possesses certain features which seem to reproduce better the physics of small  superconductors at the temperatures much below the critical one and coupled to external fields. First of all the eigenstates of the Hamiltonian (\ref{hamcpb1}) are also eigenvectors of the electric charge operator which is a well-controlled observable for small electrodes. Moreover, in contrast to the Bogoliubov-Valatin model Hamiltonian, here the states containing excited pairs are manifestly separated from the states containing unpaired electrons, not by the energy difference, but by a certain \emph{selection rule}. Indeed, the Cooper pair states are invariant with respect to time reversal operation while the states containing unpaired electrons are not. Similarly, the collective spin Hilbert space is invariant with respect to all Hamiltonian perturbations depending on Cooper pair ("spin") operators $\hat{s}^{\alpha}_k$ only. Among them there are:
\begin{enumerate}
\item Kinetic energy term (\ref{bcs1}),
\item "Local" electric potentials of the form 
\begin{equation}
\hat{U} = \sum U_k \hat{s}^{z}_k,
\label{epot}
\end{equation}
\item Cooper pair tunneling Hamiltonian from/to an external reservoir given by 
\begin{equation}
\hat{T}_{\mathrm{red}} = \frac{1}{2} \sum_{k=1}^K (\bar{\beta_k}\hat{s}^+_{k} + \beta_k\hat{s}^-_{k}),~~~\beta_k \in \mathbb{C},
\label{tunhamred}
\end{equation}
\item  Scattering of Cooper pairs or tunneling through an internal junction, governed by
\begin{equation}
\hat{H}_{\mathrm{s}} = \sum_{k,l=1}^K T[k | l]\hat{s}^+_{k}\hat{s}^-_{l}.
\label{scat}
\end{equation}
\end{enumerate}

Therefore, at low temperature regime when the number of thermally excited unpaired electrons is negligible, and under the assumption (\ref{gap3}) the physically relevant states of a small superconductor can be well approximated by the low energy sector of the considered collective spin model.

\section{Cooper pair box}
 
A Cooper pair box called "charge qubit" is a circuit consisting of a small superconducting island  with a small capacitance $C$ connected  via Josephson junction to a large superconducting reservoir. The Hamiltonian of the isolated small electrode has form (\ref{hamcpb1}) with the charging energy $E_C$ comparable to $kT_c$ and hence by (\ref{gap4}) to $g$. Notice that for a typical CPB  our assumption (\ref{gap3}) gives $K \simeq 10^4$ while the standard one (\ref{gap2}) yields $K \simeq 10^6$.

In the following we  restrict ourselves to the  energy levels corresponding to the two highest eigenvalues of $\hat{\mathbf{J}}^2$ given by
$j(j+1)$ with $j= K/2$ and $j=K/2 -1$. The corresponding eigenvectors and eigenvalues have the following form: the nondegenerate level $|K/2,m\rangle$ with the energy
\begin{equation}
  E(K/2,m) = -\frac{g(K+2)}{4} + \frac{g}{K}m(m-1)+ 4E_C (m-\bar{m})^2,
\label{ndeg}
\end{equation}
 and  the $(K-1)$-degenerate level $|K/2-1,m;r\rangle$, $r= 2,\ldots,K$, with the energy
\begin{equation}
E(K/2-1,m) = E(K/2,m)+ g.
\label{deg}
\end{equation}
Figure 2 presents the energies $E(K/2,m)$ and $E(K/2-1,m)$ for $m = -1 , 0, 1$ and $K >> 1$ as functions
of the control parameter $\bar{m}$ in terms of the energy scale $E_C$  and the dimensionless parameter $\tilde{g} = g/8E_C = 0.4$ which is close to  typical values for most of the CPB's implementations. In this limit all levels are equidistant. 

\begin{figure}[!htb]
\begin{center}
\includegraphics[width=0.8\textwidth]{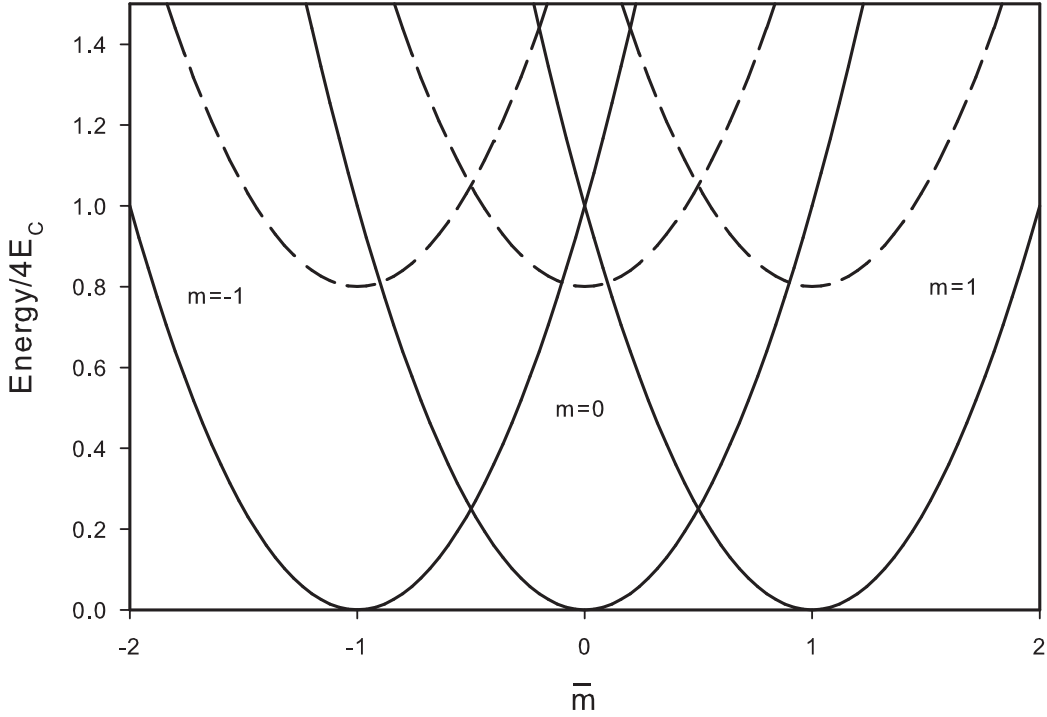}
\end{center}
\caption{The energies $E(K/2,m)+ g(K+2)/4 $ (solid line) and  $E(K/2-1,m)+ g(K+2)/4 $ (dashed line) for different values of $m$, $K >> 1$ and $\tilde{g}=0.4$ as functions
of the control parameter $\bar{m}$.}
  \label{fig1}
\end{figure}
 
Due to the \emph{Coulomb blockade} described by the second term in (\ref{hamcpb1}) one can consider only the states with $m=m_0$, $m_0 +1 $ satisfying $\bar{m}\in [m_0 , m_0 +1]$. Taking into account that $m_0 << K$ and extracting the irrelevant common constant $2E_C+4E_C [(m_0-\bar{m})^2+ (m_0-\bar{m})]- gK/4$ we have an effective Hilbert space spanned by the vectors with corresponding energies
denoted by simplified symbols
\begin{equation}
\begin{array}{ll}
|0\rangle \equiv |K/2,m_0\rangle,         &\ E_0 = -2E_C (1- 2n_g),\\
|1\rangle\equiv |K/2,m_0 +1\rangle,        &\ E_1 = 2E_C [1-2n_g - 4\tilde{g}],\\
|s;0\rangle \equiv |K/2-1,m_0;s\rangle,     &\ W_0 = -2E_C [1-2n_g - 4\tilde{g}],\\
|r;1\rangle\equiv |K/2-1,m_0+1;r\rangle,   &\ W_1 = 2E_C (1- 2n_g),
\end{array}
\label{levels}
\end{equation}
where $s,r = 2,\ldots,K$ and
\begin{equation}
n_g = \bar{m} - m_0 - \tilde{g}.
\label{levels1}
\end{equation}

\subsection{The effective Hamiltonian of CPB}
 
One should now include the coherent tunneling process through the junction between a small island and a large superconducting grounded electrode given by (\ref{tunhamred}) and treated as a small perturbation of (\ref{hamcpb1}). 
Introducing the total amplitude $\beta=  \sum_{k=1}^K \beta_k $
one can decompose the tunneling Hamiltonian into collective and individual parts
\begin{eqnarray}
\hat{T}_{\mathrm{red}}&&=\hat{T}_{\mathrm{red}}^c+\hat{T}_{\mathrm{red}}^i=\frac{1}{K}\left(\mbox{Re}(\beta)\hat{J}_{x}+ \mbox{Im}(\beta)\hat{J}_{y}\right) \nonumber\\
&&+\frac{1}{2}\sum_{k=1}^K \left[\left(\beta_k -\frac{1}{K}\beta\right)\hat{s}^+_k + \left(\bar{\beta}_k -\frac{1}{K}\bar{\beta}\right)\hat{s}^-_k\right].\nonumber\\
\label{tunham2}
\end{eqnarray}

The collective part $\hat{T}_{\mathrm{red}}^c$  preserves the subspaces of the given $j$. Therefore,   if the collective part dominates one could consider only the states with $j = K/2$ to obtain the standard  large spin model. To compare the magnitude of the collective component of the Josephson energy $E_J^c =|\beta|$ with its individual counterpart given by
\begin{equation}
E_J^i = \left[\sum_{k=1}^K \left|\beta_k -\frac{1}{K}\beta \right|^2 \right]^{1/2} =\left[\sum_{k=1}^K |\beta_k|^2 -\frac{1}{K}|\beta |^2 \right]^{1/2}
\label{ind}
\end{equation}
one can consider a simple toy model with $\beta_k = Ae^{i\lambda k}$, $k= 0,1,\ldots,K-1$. Then
\begin{eqnarray}
E_J^c& =& |A|\frac{|1-e^{i\lambda K}|}{|1-e^{i\lambda}|}\leq \frac{2|A|}{|1-e^{i\lambda}|},\nonumber\\
E_J^i& \simeq & |A|\sqrt{K}
\label{ind1}
\end{eqnarray}
what implies for a generic $\lambda$ that $E_J^i \sim E_J^c \sqrt{K}$.

On the other hand for purely random amplitudes $\beta_k$,  $|\beta|^2 = \sum_k |\beta_k|^2$ and therefore $E_J^i \simeq E_J^c$. The real system should be placed between these two extremal cases of strong interference and random behavior what implies that the ratio $E_J^i / E_J^c $
increases as a certain positive power of $K$ leading to the domination of the individual coupling.
For large JJs with small $E_C$ this effect is suppressed by the fact that the level splitting for a fixed $j$ determined by the Coulomb repulsion is much smaller than the level splitting for different values of $j$ given by the superconducting gap. For small junctions those energy scales are comparable and the individual tunneling prevails. This implies
that the matrix elements of $\hat{T}_{\mathrm{red}}$ between the vectors with the same $j$ are negligible in comparison
with the elements between vectors with $|j-j'|=1$. Hence the only relevant matrix elements are the following
\begin{eqnarray}
\langle r;1|\hat{T}_{\mathrm{red}}|0\rangle & =& \langle r;1|\hat{T}^{i}_{\mathrm{red}}|0\rangle = \eta_r,\nonumber\\
\langle s;0|\hat{T}_{\mathrm{red}}|1\rangle &=& \langle s;0|\hat{T}^{i}_{\mathrm{red}}|1\rangle = {\eta'}_{s}.
\label{tunhamred2}
\end{eqnarray}
Since the states $|K/2, m\rangle$ are totally symmetric the matrix elements  $\langle 1|\hat{s}^+_k|0\rangle$ ($k=1,\ldots,K$) are  equal to a constant independent of the index $k$  and it is easy to show that $\langle 1|\hat{T}^{i}_{\mathrm{red}}|0\rangle =0$.
 
We can write down the full effective Hamiltonian of the CPB including  (\ref{hamcpb1}) and (\ref{tunhamred2}) which is a direct sum of two similar terms
\begin{equation}
\hat{H}_{CPB}= \hat{H}^0_{CPB}\oplus\hat{H}^1_{CPB}
\label{sumham}
\end{equation}
acting on the subspaces $\mathcal{H}^0_{\mathrm{eff}}$ and $\mathcal{H}^1_{\mathrm{eff}}$ spanned by $\{|0\rangle , |r;1\rangle , r = 2,\ldots,K\}$ and
$\{|1\rangle , |s;0\rangle , s = 2,\ldots,K\}$, respectively. To a large extend both "subsystems" can be treated separately and completely analogically. Therefore, in the following we restrict ourselves to the first one with the Hamiltonian (comp. (\ref{levels}))
\begin{eqnarray}
\hat{H}^0_{CPB}&=& \frac{E(n_g)}{2}\left(|\xi\rangle\langle\xi|-|0\rangle\langle 0|\right)
  + \frac{E_J}{2} \left(|0\rangle\langle \xi|+ |\xi\rangle\langle 0| \right) \nonumber \\
  &&+\frac{E(n_g)}{2} \hat{P}_0. 
\label{hamcpb2}
\end{eqnarray}
Here
\begin{equation}
  E(n_g) = 4E_C (1-2n_g),~~~E_J = 2\left(\sum_{r=1}^{K-1}|\eta_r|^2\right)^{1/2},
\label{hamcpb3}
\end{equation}
\begin{equation}
|\xi\rangle =\sum_{r=2}^{K}\xi_r|r;1\rangle,~~~\hat{P}_0 = \sum_{r=2}^{K}|r;1\rangle\langle r;1| - |\xi\rangle\langle\xi|
\label{hamcpb4}
\end{equation}
with $\xi_r = 2\eta_r/E_J$.
The parameter $E_J$ is the Josephson energy describing the transition rate from $|0\rangle$ to the state given by a normalized vector $|\xi\rangle$. The value of $E_J$ can be controlled by external magnetic field and typically $E_J < E_C$.
 
Introducing two vectors $|\pm\rangle$
\begin{eqnarray}
|+\rangle & =& \cos\frac{\theta}{2}|\xi\rangle +\sin\frac{\theta}{2}|0\rangle, \nonumber\\
|-\rangle &=& \cos\frac{\theta}{2}|0\rangle -\sin\frac{\theta}{2}|\xi\rangle,
\label{eigenvec}
\end{eqnarray}
where $\theta$ is defined by $\cos\theta=E(n_g)/\sqrt{E(n_g)^2 + E_J^2}$, and the \emph{qubit observables}
\begin{eqnarray}
\hat{\sigma}^+&=&\frac{1}{2}\left(\hat{\sigma}^x + i\hat{\sigma}^y\right) = |+\rangle\langle -|,\nonumber\\
\hat{\sigma}^z &=& |+\rangle\langle +| - |-\rangle\langle -|, \label{qubit}\\
\hat{\sigma}^0& =&|+\rangle\langle +| + |-\rangle\langle -|\nonumber
\end{eqnarray}
we obtain from (\ref{hamcpb2})  a new form of the Hamiltonian
\begin{equation}
\hat{H}^0_{\mathrm{CPB}} =  \frac{1}{2}\left(\omega (n_g) \hat{\sigma}^z  + E(n_g)\hat{P}_0\right)
\label{qubitham}
\end{equation}
with two eigenvectors $|+\rangle$ and $|-\rangle$ separated by the energy difference 
\begin{equation}
\omega (n_g)=\sqrt{[4E_C(1- 2n_g)]^2 + E_J^2} 
\label{omCPB}
\end{equation}
and the third $(K-2)$-fold degenerate level corresponding to $\hat{P}_0$ with the energy $E(n_g)/2$.
This third level lies always between  $|+\rangle$ and $|-\rangle$ as $-\omega/2 \leq E(n_g)/2 \leq \omega/2$.
The relevant energy levels as functions of $n_g$  are showed in Fig. 2. 

\begin{figure}[!htb]
\begin{center}
\includegraphics[width=0.8\textwidth]{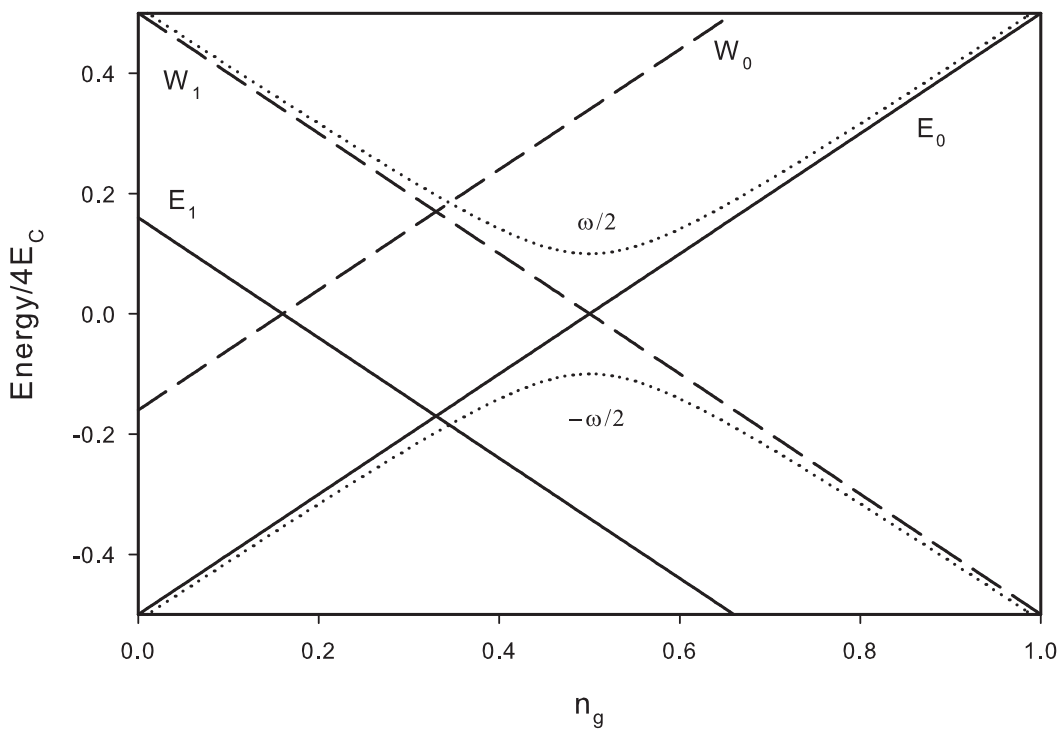}
\end{center}
\caption{The energies of the states $|+\rangle$ and $|-\rangle$ (dotted lines) and the energies given by (\ref{levels})  as functions of $n_g$ for $\tilde{g} = 0.17$ and $\delta = E_J/4E_C = 0.2$. }
  \label{fig2}
\end{figure}

The external control is performed by changing $n_g$ and applying a microwave radiation.
Obviously, if the system is completely isolated the \emph{qubit space} spanned by $|+\rangle$ and $|-\rangle$ is invariant with respect to the Hamiltonian and the external control yielding the usual model of  charge qubit.
The third, $(K-2)$-fold degenerate, energy level corresponding to $\hat{P}_0$ and the lowest energy levels of the second subsystem become important when the coupling to an environment is taken into account.

\section{Flux qubit}

A flux qubit (FQ) is a small superconducting ring interrupted by  one or several Josephson junctions. The main difference between this system and the CPB is the existence of the additional quantum number $\mu$ which accounts for the quantized circular motion of the Cooper pair as a whole. Consider first a collective spin model of such a ring without junction treated as a system of $K\times(2 \mu_{max}+1)$ spins-$1/2$ with spin operators $\hat{s}^{\alpha}_{k\mu}$, $\alpha = x,y,z$, satisfying
\begin{equation}
[\hat{s}^x_{k\mu} ,\hat{s}^y_{l\nu}] = i\delta_{kl}\delta_{\mu\nu}\hat{s}^z_{k\mu}
\label{spin}
\end{equation}
and cyclic permutations of $x,y$ and $z$. Here, the index $k = 1,2,...K/2$ corresponds to the internal quantum numbers characterizing a Cooper pair in its center of motion reference frame. The second quantum number $\mu= 0, \pm 1, \pm 2,...,\pm \mu_{max}$ labels  quantized circular states for the Cooper pairs center of motion with respect to the center of the ring.
The maximal accessible value of $|\mu|$ is a consequence of the maximal critical current $I_0$ which can flow in the ring at zero temperature. This effect can be described by $\mu$-dependence of the coupling constant $g_{\mu}$ due to the fact that the states with a nonzero angular momenta generate current $I$ proportional to $|\mu|$ which modifies the gap according to the formula
\begin{eqnarray}
g(I)&=&g \left[1- \left(\frac{I}{I_0}\right)^2\right]
\label{gapI}
\end{eqnarray}
or
\begin{eqnarray}
g_{\mu}&=&g \left[1- \left(\frac{\mu}{\mu_{max}+1}\right)^2\right].
\label{gapIa}
\end{eqnarray}
To justify (\ref{gapI}) one should notice that  $-g/2$ is the pairing energy of a single Cooper pair in the center of mass reference frame. When the Cooper pair moves with a velocity $v$, producing a current density $ I\sim v$, its energy increases  to  $-g/2 + \mathrm{const} \times I^2 \equiv - g(I)/2$ due to its kinetic energy and the energy of the created magnetic field. When the current density reaches the value $I_0$, such that the effective pairing energy $g(I_0)= 0$, it is favorable for the system to relax from the superconducting state with a current $I_0$ to a normal state with zero current and higher entropy. 

Defining again the collective spin  operators $\hat{\mathbf{J}}_{\mu}= (\hat{J}^x_{\mu} ,\hat{J}^y_{\mu} ,\hat{J}^z_{\mu} )$ by
\begin{equation}
\hat{J}^{\alpha}_{\mu}= \sum_{k=1}^K\hat{s}^{\alpha}_{k\mu},~~~\alpha = x,y,z,
\label{spincoll}
\end{equation}
and Cooper pair number operators 
\begin{equation}
\hat{N}_{\mu}= \sum_{k=1}^K \left(\hat{s}^{z}_{k\mu} +\frac{1}{2}\right)= \hat{J}^z_{\mu} +\frac{K}{2}
\label{number}
\end{equation}
one can write the strong coupling limit BCS Hamiltonian for a ring as
\begin{eqnarray}
\hat{H}_{\mathrm{ring}}&=& -\frac{1}{K}\sum_{\mu} g_{\mu}\left(\hat{\mathbf{J}}_{\mu}^2 - (\hat{J}^z_{\mu})^2 + \hat{J}^z_{\mu}\right)\nonumber\\
&&+ E_L \left(\sum_{\mu}\mu \hat{N}_{\mu} - \mu_{ext}\right)^2.
\label{hamFQ} 
\end{eqnarray}
The first term in the Hamiltonian (\ref{hamFQ}) is the standard mean-field BCS pairing Hamiltonian. 
The second term is also of the mean-field type what is the reasonable approximation for the rings with a thickness not larger than the penetration depth for a magnetic field (typically $\sim 100$ nm). In these cases we can ignore spatial variations of the current and  derive (\ref{hamFQ}) using the macroscopic relation between the current density $\mathbf {j}$ and the vector potential  $\mathbf{A}$ in a superconductor\cite{Fey}
\begin{equation}
\mathbf j = - \frac{\mathcal{N}e^2}{m}\mathbf{A} 
\label{jvsA}
\end{equation}
and the energy of the current in a magnetic field
\begin{equation}
\mathcal{E}= -\int \mathbf {j}\cdot\mathbf{A}\, d^3\mathbf{x} = \frac{K e^2}{m} A^2 .
\label{Emag}
\end{equation}
Here $\mathcal{N}$ is a density of superconducting electrons, $m$ is an electron mass and $A = |\mathbf{A}|$ is assumed constant along the loop. Using now the quantization condition for the magnetic flux that threads the loop of the length $\ell$
\begin{equation}
A\cdot\ell = F \Phi_0,~~~~ F= 0,\pm 1,\pm 2,\ldots,
\label{Emag1}
\end{equation}
where $\Phi_0 = h/2e$, one obtains the quantized energy
\begin{equation}
\mathcal{E}(F)= \frac{K h^2}{4m\ell^2} F^2 .
\label{Emag2}
\end{equation}
Replacing $F$ by the flux operator $\hat{F} = \sum_{\mu}\mu \hat{N}_{\mu}$ and adding a shift $\mu_{ext} = \Phi_{ext}/\Phi_0$ caused by the external magnetic flux $\Phi_{ext}$ one obtains the second term in (\ref{hamFQ}) with 
the inductive energy given by
\begin{equation}
E_L= Kh^2/4m\ell^2.
\label{EL}
\end{equation}
The above choice of flux operator is consistent with the Onsager hypothesis that the flux generated by the circulating charge $2e$ is quantized in the units of $\Phi_0$\cite{Ons}.

Notice that  we neglect here the Coulomb repulsion term as the charging energy $E_C$ is much smaller than for the CPB and the number of Cooper pairs in the system is fixed.

The physical Hilbert space is spanned by the joint eigenvectors  
\begin{eqnarray}
\hat{\mathbf{J}}_{\mu}^2|...(j_{\mu};n_{\mu};r_{\mu})...\rangle& =& j_{\mu}(j_{\mu}+1)|...(j_{\mu};n_{\mu};r_{\mu})...\rangle,\nonumber\\
\hat{N}_{\mu}|...(j_{\mu};n_{\mu};r_{\mu})...\rangle& =& n_{\mu}|...(j_{\mu};n_{\mu};r_{\mu})...\rangle
\label{jeigen}
\end{eqnarray}
with the multiplicity $r_{\mu}$ and satisfying the condition
\begin{equation}
\sum_{\mu} n_{\mu} = \frac{K}{2}
\label{norm}
\end{equation}
which determines the total number of Cooper pairs in the system.

\subsection{Ground state and lowest excitations}
Consider first the case $\mu_{ext} = 0$. The unique ground state of the Hamiltonian (\ref{hamFQ}) can be obtained by minimizing the energy given by the first term. Indeed, due to the symmetry $g_{\mu} = g_{-\mu}$ the contribution from the second term automatically vanishes for such minimizers. One can easily show that the ground state has a product structure
\begin{equation}
|0\rangle = |...(K/2;\tilde{n}_{\mu};1)...\rangle,           
\label{groundFQ}
\end{equation}
where $\tilde{n}_{\mu} \simeq \frac{K}{2}p_{\mu}$ and the ocupation probabilities $p_{\mu}$ minimize the functional $\sum_{\mu} g_{\mu}(p_{\mu}^2 - 2p_{\mu})$. The numerically obtained shape of the probability distribution $p_{\mu}$ is presented in Fig. 3.

\begin{figure}[!htb]
\begin{center}
\includegraphics[width=0.8\textwidth]{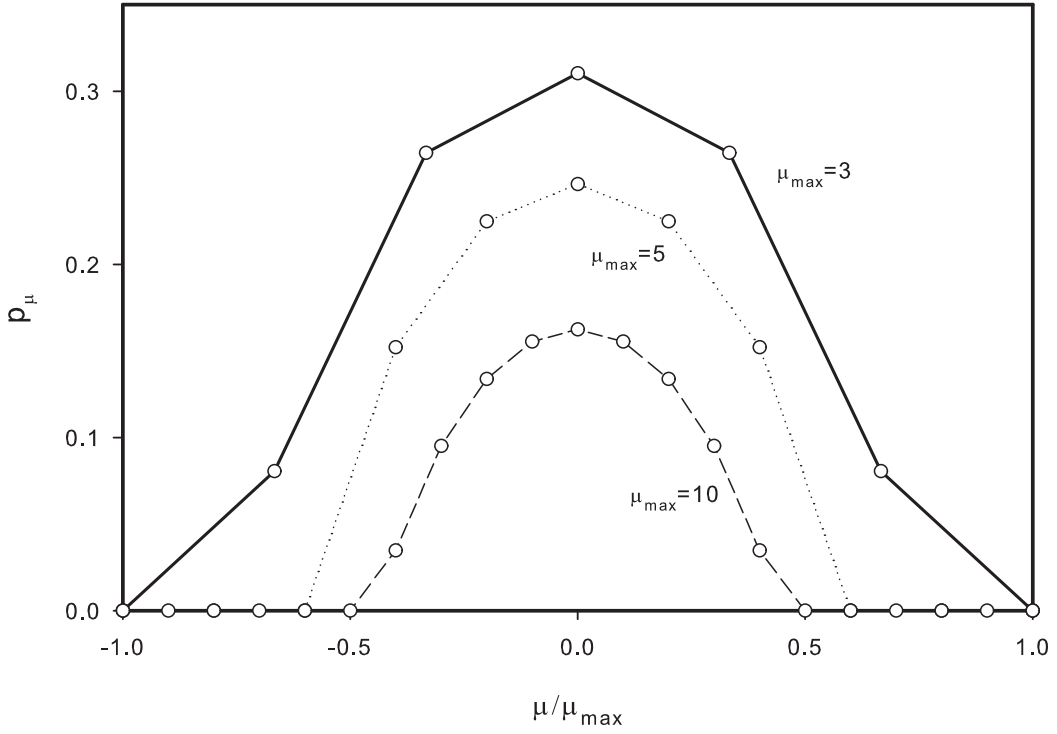}
\end{center}
\caption{Ground state occupation probabilities of different circular states. Notice that probabilities vanish far away from the borders ensuring stability of the ground state. }
  \label{fig3}
\end{figure}

The spectral decomposition of the Hamiltonian (\ref{hamFQ}) is very rich and contains different types of manifolds of the lowest lying excited states. The first type is obtained by the creation of a single excited pair without changing the zero magnetic flux of the ground state, the second one is spanned by the vectors with the flux quantum number $F=\pm 1$  and the third corresponds to both types of excitations. The external  magnetic flux $\Phi_{ext}= \mu_{ext}\Phi_0$, $\mu_{ext}\simeq 1/2$,  shrinks the gap between the ground state and the states with $F =1$ what is necessary to separate two qubit states from the rest. On the other hand, similarly to the case of CPB individual tunneling processes create excited pairs. Therefore, to construct a proper model of FQ we need only the third type of excited states (flux $F=1$ and one excited pair) represented by the vectors
%
\begin{widetext}
\begin{equation}
|\nu;r\rangle = |(K/2;\tilde{n}_{-\mu_{max}};1)\ldots(K/2;\tilde{n}_{\nu}-1;1)(K/2-1;\tilde{n}_{\nu+1}+1;r)\ldots(K/2;\tilde{n}_{\mu_{max}};1)\rangle 
\label{excFQ}
\end{equation}
\end{widetext}
with $\nu = -\mu_{max},-\mu_{max}+1,...,\mu_{max}$ and $r =2,3,...,K$.

\subsection{The FQ Hamiltonian}
For small rings the condition $E_L \gg g$ holds and therefore, in order to produce a qubit, we have to switch on the external magnetic flux $\Phi_{ext}\simeq \Phi_0/2$, ($\mu_{ext}\simeq 1/2$). Then the gap between the ground state and the states (\ref{excFQ}) is given by
\begin{equation}
E_{0\nu} =  E_L (1- 2\mu_{ext}) + g_{\nu +1}.
\label{gapFQ}
\end{equation}
The presence of the junction(s) can be modeled by the generalization of the Hamiltonian (\ref{scat})
\begin{equation}
\hat{H}_{\mathrm{s}} = \sum_{\mu\neq\nu} \sum_{k,l=1}^K T[k,\mu | l, \nu]\hat{s}^+_{k\mu}\hat{s}^-_{l\nu},   
\label{scham}
\end{equation}
where $T[k,\mu | l, \nu]$ is a hermitian matrix of scattering amplitudes.

Again, similarly to the CPB case, the collective scattering processes which preserve quantum numbers $j_{\mu}$ can be neglected in comparison with the individual scattering changing $j_{\mu}$.
Therefore, the relevant matrix elements of the perturbation (\ref{scham}) are the following
\begin{equation}
\langle \nu; r|\hat{H}_{\mathrm{s}}|0\rangle  = \eta_{\nu r}.
\label{scatt}
\end{equation}
Similarly, to the CPB case we can write the effective Hamiltonian of the first qubit (the second one is initialized by decreasing $\mu_{ext}$ to  $1/2$) in the lowest order perturbation and projected on the the 2-dimensional qubit space
\begin{equation}
\hat{H}_{\mathrm{FQ}}= \frac{1}{2}\left(E(\tilde\mu_{ext})\left(|\xi\rangle\langle\xi|-|0\rangle\langle 0|\right)
  + E_J \left(|0\rangle\langle \xi|+ |\xi\rangle\langle 0| \right)\right).
\label{Fqubit}
\end{equation}
Here
\begin{eqnarray}
E(\tilde\mu_{ext})& =& E_L (1-2\tilde\mu_{ext}),\label{Fqubit1}\\
\tilde\mu_{ext}&=& \mu_{ext}- \frac{1}{2E_L}\sum_{\nu,r}|\xi_{\nu r}|^2 g_{\nu +1},\\
E_J& =& 2\left(\sum_{r=2}^{K}\sum_{\nu}|\eta_{\nu r}|^2\right)^{1/2},
\label{Fqubit2}
\end{eqnarray}
\begin{equation}
|\xi\rangle =\sum_{r=2}^{K}\sum_{\nu}\xi_{\nu r}|\nu;r\rangle,~~~\xi_{\nu r} = 2\eta_{\nu r}/E_J.
\label{Fqubit3}
\end{equation}
In the above we omit the part of the Hamiltonian which describes the "sink states" decoupled from the qubit states.
Again we observe the structure of levels  qualitatively the same as for the CPB (see Fig. 2) with characteristic qubit level repulsion caused by a tunneling.

The final form of the FQ Hamiltonian reads (compare with (\ref{qubitham}-\ref{omCPB}))
\begin{eqnarray}
\hat{H}_{\mathrm{FQ}}& =&  \frac{1}{2}\omega(\tilde\mu_{ext}) \hat{\sigma}^z,\nonumber\\
\omega(\tilde\mu_{ext})&=&\sqrt{[E_L(1- 2\tilde\mu_{ext})]^2 + E_J^2}. 
\label{Fqubit4}
\end{eqnarray}
The qubit is controlled by changing the external flux $\tilde\mu_{ext}\Phi_0$ and applying microwave radiation.
\section{Current-biased junction}
This type of JJ, denoted by CBJ, consists of larger superconducting electrodes than in the CPB device what implies that the Coulomb energy $E_C$ is much smaller than the gap parameter $g$. The electrodes are connected to a current source which produces a constant but tunable current $I$. We propose a microscopic model of CBJ using again the reduced BCS Hamiltonian (\ref{bcs1}) and emphasizing the role of excited Cooper pairs. In the first approximation we treat the system as a closed superconducting circuit with a low value of $E_L$ which can support a steady current $I$ and the junction acting as a perturbation -- a scattering center. The starting point is the effective unperturbed Hamiltonian similar to (\ref{hamFQ}) under the assumption that $E_L \ll g(I)$ (see (\ref{gapI})).
The charging energy (neglected in (\ref{hamFQ})) is also small ($E_C \ll g(I)$) and the number of Cooper pairs is not fixed what implies that both quantum numbers $\mu$ and $n_{\mu}$ correspond to  certain essentially classical degrees of freedom. The collective coupling of these degrees of freedom  to an environment produces, by mechanisms mentioned in Sec. I,  semiclassical coherent-like states  determined by the external conditions. Therefore, the ground state of the system can be written as (compare the structure of eigenvectors (\ref{jeigen}))
\begin{widetext}
\begin{equation}
|0\rangle = \sum_{\mu}\sum_{n} \phi_I (\mu)\psi_K (n)|...(j_{\mu}= K/2;n_{\mu}=n ; r\equiv 1)...\rangle,
\label{refst}
\end{equation}
\end{widetext}
where the probability amplitudes $\phi_I (\mu)$ and $\psi_K (n)$ display normal fluctuations around
mean values $\mu_I \sim I $ and $K/2$, respectively. In order to construct the lowest excited states we apply a kind of adiabatic approximation fixing the semi-classical degrees of freedom and changing the only quantum one related to a number of excited pairs $p$. Then the structure of  strongly degenerate excited states is the following
\begin{widetext}
\begin{equation}
|r_p ; p\rangle = \sum_{\mu}\sum_{n} \phi_I (\mu)\psi_K (n)|...(j_{\mu}= K/2-p;n_{\mu}=n; r_p)...\rangle,
\label{cohpairs}
\end{equation}
\end{widetext}
where $p \ll K$ and $r_p$ describes degeneracy. 

\emph{Remark:} The derivation of above is valid for a single JJ in the phase qubit regime. If the device is designed as a loop, the loops inductive energy should be taken into account. The additional quantized energy produces different
initial flux states replacing the ground state (\ref{refst}). They yield the different critical  currents  as the internal loop current adds to the external biased one. Therefore, the initialization of the proper flux state must be done before the device can be used as a qubit\cite{Pal}.

Similarly to CPB and FQ we consider a qubit model including only the ground state $|0\rangle$
and the $(K-1)$-degenerate first excited states $|r;1\rangle$ separated by the energy gap $g(I)$. Again the junction acts as a scattering center given by the Hamiltonian (\ref{scham}) which couples the ground and excited states as in (\ref{scatt}). Repeating the analogical construction we obtain the following qubit Hamiltonian
\begin{equation}
\hat{H}_{\mathrm{CBJ}} =  \frac{1}{2}\left(\omega(I) \hat{\sigma}^z  + g(I)\hat{P}_0 \right),
\label{qubitham1}
\end{equation}
where 
\begin{equation}
\omega(I) =\sqrt{g(I)^2 + E_J(I)^2}.  
\label{omI}
\end{equation}
The two eigenvectors $|+\rangle$ and $|-\rangle$ of (\ref{qubitham1}) are separated by the energy difference $\omega(I)$ an the
third $(K-2)$-fold degenerate level with the energy $g(I)/2$ corresponds to $\hat{P}_0$.
This third level lies always between  $|+\rangle$ and $|-\rangle$, as $-\omega/2 \leq g/2 \leq \omega/2$ (Fig. 4).

\begin{figure}[!htb]
\begin{center}
\includegraphics[width=0.8\textwidth]{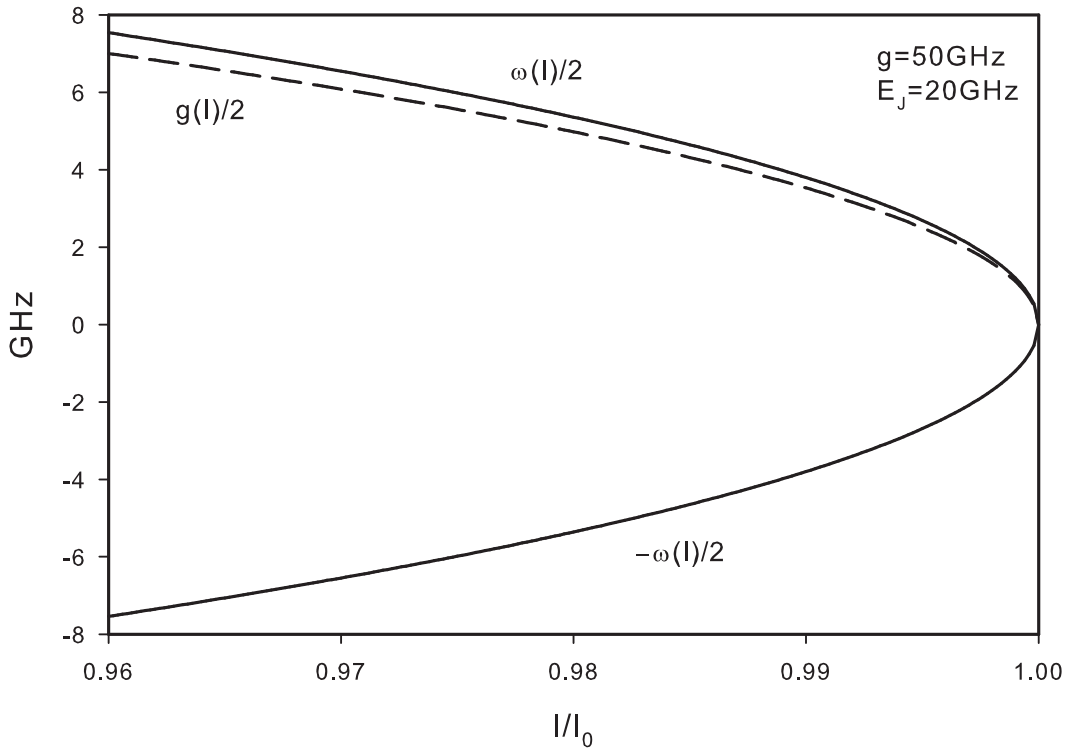}
\end{center}
\caption{Qubit levels (solid line) and the sink level (dashed line) as functions of biased current $I$. }
  \label{fig4}
\end{figure}

Similarly to the previous cases the external control is performed by  tunning the biased current and  the coupling  to microwave radiation.

\section{Comparison with standard theory and experiments} 

The main predictions of the proposed unified  microscopic model of small JJs concern the structure and parametrization of their lowest energy levels. The comparison of our model with standard theories and experimental data is presented below for all three
types of superconducting qubits. There are also another aspects of our approach related to dissipation/decoherence processes which will be discussed in the forthcoming paper.
\subsection{Charge qubits}
The Kirchhoff equation for the CPB is equivalent to the Hamilton equations obtained from the following classical Hamiltonian of a fictitious particle\cite{W} 
\begin{equation}
H = \frac{(p - p_g)^2}{2m} - E_J \cos{x}. 
\label{chargeham}
\end{equation}
Here the "position" $x$ corresponds to the phase variable, $p_g \sim n_g$ is a control parameter and the "mass" is given by
\begin{equation}
m=\left(\frac{\hbar}{2e}\right)^2 C . 
\label{mass}
\end{equation}
The harmonic approximation around the potential minimum
gives the value of the plasma freqency 
\begin{equation}
\omega_{p} =\sqrt{\frac{4\pi e I_0}{\hbar\, C}}.
\label{plasma}
\end{equation}
However, for small CPBs one does not use a quantized version of (\ref{chargeham}) but rather a two-level approximation
to the large spin Hamiltonian (\ref{cpbham1})
\begin{equation}
\hat{H} = -\frac{1}{2}\left[ E_C(1-2n_g)\hat{\sigma}^z  + E_J \hat{\sigma}^x\right]. 
\label{chqubit} 
\end{equation}
The Hamiltonian (\ref{chqubit}) is essentially the same as ours (\ref{hamcpb2}) and predicts the same  qubit frequency as a function of $E_C$ and $E_J$. Therefore, to test our model one should explore the differences. The consequences of the presence of the highly degenerate level (probability sink) for the decoherence processes will be discussed in the forthcoming paper. Here we refer to Fig. 2 and the formulas (\ref{levels}) which show that when the control parameter $n_g$ is ramped from zero to the value $1/2$ the ground state of the initialized qubit is unstable due to the transitions to the level $E_1$ in the range $(1/2 - \tilde{g})\leq n_g \leq 1/2$. 
A similar instability is predicted in the range $1/2\leq n_g \leq1/2 + \tilde{g}$ for the case of $n_g$ decreasing from $1$ to $0$.
 
The instability of the ground state for $0.3 < n_g < 0.7$ was reported by Lehnert \emph{et al.}\cite{Leh} and attributed to backaction generated by currents flowing by the RF-SET device which  is used to measure the charge of the CPB. Here we give an intrinsic explanation of this effect supported by the data. Indeed, taking experimental\cite{Leh} value $4E_C/h = 149$ GHz and puting $g/k_B = 2T_c = 2.4$ K for Al, one obtains $\tilde{g}= 0.17$ what agrees very well with the observed instability range.

\subsection{Flux qubits}

The standard description\cite{W} is based on the quantization of the following modification of (\ref{chargeham})
\begin{equation}
H = \frac{p^2}{2m} - E_J \cos{x} + E_L \left(\frac{x- x_{ext}}{2\pi}\right)^2,  
\label{fluxham}
\end{equation}
where $x = 2\pi\Phi/\Phi_0$ and $x_{ext} = 2\pi\Phi_{ext}/\Phi_0$. Here $\Phi$ is  a magnetic flux that threads the loop and $\Phi_{ext}$ is the external flux applied to the loop. In the standard approach one has to compute the spectrum of the quantized version of (\ref{fluxham}) using the quartic approximation which leads to a double well potential picture. One should notice that the frequency corresponding to a two level approximation depends on three parameters characterizing the device $E_C$, $E_J$, $E_L$ and one control parameter $\Phi_{ext}$. Applying the prediction of our model (\ref{Fqubit4}) under the assumption $|E_L(1- 2\tilde\mu_{ext})|>> E_J$ we obtain a very simple formula for the qubit frequency $f$
\begin{equation}
h f \simeq |E_L(1- 2\tilde\mu_{ext})|
\label{fluxf}
\end{equation}
depending on the microscopic parameters  and the geometry of the sample (see (\ref{EL})).

This result can be compared with the experimental data\cite{Wal}. The linear dependence given by (\ref{fluxf}) far enough from $\tilde\mu_{ext} = 1/2$ is clearly confirmed and the value $E_L/h \simeq 1.5 \times 10^3$ GHz can be extracted from the Fig. 3.(B) presented by van der Wal \emph{et al.}\cite{Wal} The sample is an aluminum $5~\mu$m $\times$
5~$\mu$m loop made of 450 nm wide and 80 nm thick lines. Using the value $\ell=20\mu$ m and the formula (\ref{EL}) we obtain the number of superconducting electrons $K = 3.3 \times 10^6$. This value can be put into the 
formula 
\begin{equation}
K = 2\hbar \omega_{cut} N(0) =2\hbar \omega_{cut} V\frac{m}{2\pi^2 \hbar^2}\left(3\pi^2 \kappa\right)^{1/3},
\label{Kvalue}
\end{equation}
where $V$ is a volume of the sample and $\kappa$ is a density of electrons ($\kappa =18.06 \times 10^{22}/$cm$^3$ for Al). The substitution yields $\hbar \omega_{cut}/2k_B = 1.3$ K which is close to the critical temperature $T_c = 1.2$ K. This is a strong support for the basic assumption of our model $ \hbar \omega_{cut} = g = 2k_B T_c$ (see (\ref{gap3})) and gives the first estimation of the zero temperature density of Cooper pairs  $K/2V = 2.5 \times 10^{18}$/cm$^3$ in a superconductor obtained directly from the experimental data.

\subsection{Phase qubits}
For this type of junction one uses a model equivalent to the fictitious particle moving in a washboard potential
with the Hamiltonian\cite{W} being again a modification of (\ref{chargeham})
\begin{equation}
H = \frac{p^2}{2m} - E_J \left(\cos{x} + \frac{I}{I_0}x\right).
\label{hamphase}
\end{equation}
The harmonic approximation around the potential minimum
gives the value of the plasma frequency 
\begin{equation}
\omega_{p} =\sqrt{\frac{4\pi e I_0}{\hbar\, C}}\left[1- \left(\frac{I}{I_0}\right)^2\right]^{1/4}  
\label{washboard2}
\end{equation}
which can be treated as a rough approximation to the qubit frequency\cite{W}. More complicated formulas which take into account anharmonicity  and involve $E_J$ as an additional parameter are also available. On the other hand our formula for the phase qubit frequency obtained by combining (\ref{gapI}) with (\ref{omI}) and under the assumption $g(I)\gg E_J$ reads
\begin{equation}
\omega(I) \simeq g \left[1- \left(\frac{I}{I_0}\right)^2\right].  
\label{phase}
\end{equation}
It is important that the formula (\ref{phase}) involves the microscopic parameter $g$ while the standard expressions depend entirely of the macroscopic ones. 

Firstly, one can check roughly the magnitude of the predicted qubit frequencies $\omega(I)$. Typically, $E_J \ll g\simeq 49$ GHz for Al and  $g\simeq 388$ GHz for Nb. In all experiments the biased current satisfies $0.85 \leq I/I_0 \leq 0.99$\cite{} what leads, using (\ref{phase}) to a reasonable range of  frequencies 0.98 GHz $\leq \omega/2\pi \leq$ 13.6 GHz for Al and
7.7 GHz $\leq \omega/2\pi \leq$  108 GHz for Nb, respectively.

More detailed comparison is performed for the several examples from the literature. One should notice that comparing the experimental data with any theoretical curve describing $\omega(I)$ which depends on free parameters is difficult and inacurate because the range of the biased current variation is in all experiments a very small fraction of the whole interval $[0, I_0]$ (see Fig. 5).

\begin{figure}[!htb]
\begin{center}
\includegraphics[width=0.8\textwidth]{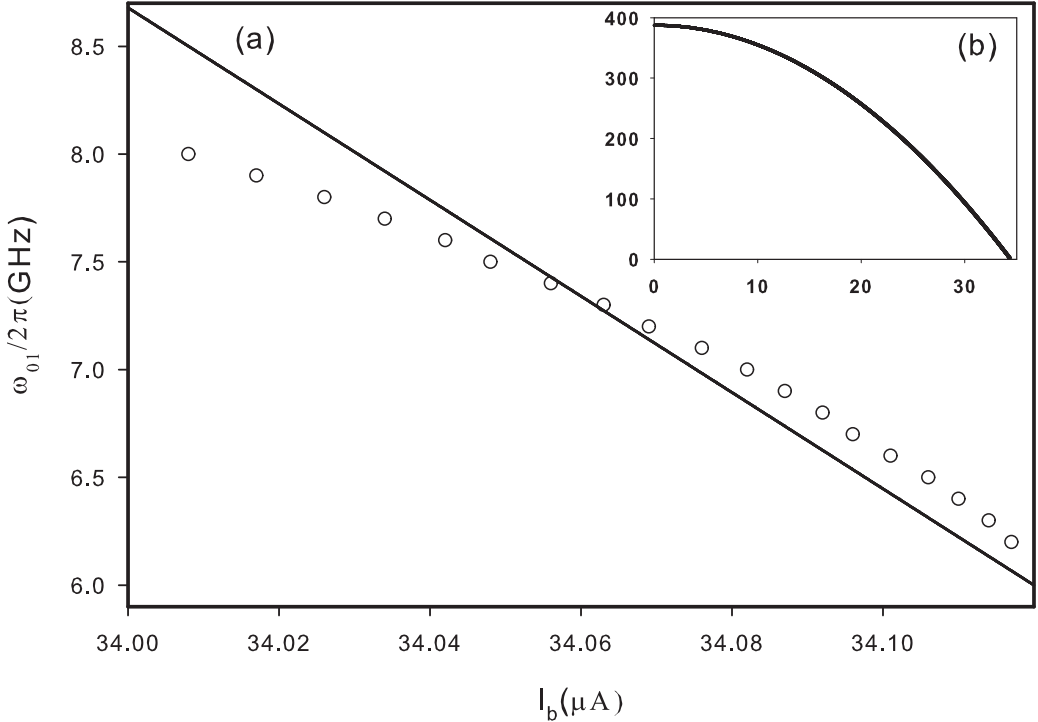}
\end{center}
\caption{ Spectrum of SQUID DS$_1$\cite{Dut} (Nb). (a) The circles represent the experimental data\cite{Dut} and the line is  fitted according to (\ref{phase}) with $I_0=34.387\mu$A ($g=2k_B T_c\simeq 388$ GHz). b) The fit in the  whole range of variation of $\omega(I)/2\pi [GHz]$.}
  \label{fig5}
\end{figure}

Table I presents the obtained values of  $I_0$ using the formula (\ref{phase}) and the experimental values of $\omega(I)/2\pi$ for few examples of phase qubits. The theoretical values are compared  with the reported experimental values of the critical current or in the case SQUID DS$_1$\cite{Dut} with the value fited to the standard theory.
 
\begin{table}[!ht]
\begin{center}
\caption{The critical currents for phase qubits realizations}
\begin{tabular}{p{3.5cm}p{2.0cm}p{2.5cm}}
\hline
\hline
Junction & $I_0$ $\mu$A (exp.)  & $I_0$ $\mu$A (theory)\\
\hline 
 SQUID DS$_1$ (Nb)\cite{Dut}  & 34.275  (fit) & 34.387 (fit) \\
 current-biased (Al)\cite{Anderson02}  & 13.33   & 13.99  \\
current-biased (Al)\cite{Mar}  & 21   & 22.4  \\
 SQUID DS$_{3A}$ (Al)\cite{Pal}  & 1.26   & 1.27  \\
\hline
\end{tabular}
\end{center}
\end{table}  

One should remember that the measurements of the critical current are not very precise because this parameter can vary in time for the same sample. The positions of experimental values of $\omega (I)$ are also quite sensitive to the level repulsion phenomena caused by "parasite" two-level systems present in the environment and interacting with a qubit\cite{Pal,Dut}.
\section{Conclusions}
The presented approach to superconducting qubits differs from the standard one by referring to the microscopic Hamiltonian being a simplified version of the BCS Hamiltonian and avoiding a detour via \emph{requantization} of the macroscopic Kirchhoff's equations. As a consequence, in contrast to the standard results, the obtained formulas describing the energy spectra depend on the microscopic parameters.  Our choice of the microscopic Hamiltonian implies also a nonstandard assumption about the cut-off energy scale
of the BCS model. We follow the original approach where this energy is of the order of superconducting gap, while in the modern literature one chooses the Debye energy which is larger by two orders of magnitude. The computation of the number of superconducting electrons in the flux qubit device, based on the experimental data, strongly supports this choice. A number of experimental results concerning the energy spectra of different types of superconducting qubits is consistent with our model as well.

 Another feature of our model is the importance of individual scattering/tunneling of excited Cooper pairs which differs from the standard picture of independent quasiparticles in the Bogoliubov-Valatin scheme. In particular the qubit states in our model are spanned by the ground state and the given single excited Cooper pair state. This is conceptually very different from the standard picture of \emph{macroscopic quantum systems} but on the other hand solves the puzzle of missing environmental effects which should produce semiclassical behavior.
Last but not least, the presented model allows new mechanisms of decoherence  due to the excited Cooper pair coupling to phonons and existence of probability sinks. Those phenomena will be studied in the forthcoming paper.
 
\begin{acknowledgments}
 R. A. is supported by the Polish Ministry of Science and Higher Education, grant PB/2082/B/H03/2010/38.
\end{acknowledgments}


\begin{thebibliography}{99}
\bibitem{g}
In the standard notation used in the literature\cite{BCS,Tin} $g$ is replaced by $\hbar\omega_{cut}N(0)V$.
\end{thebibliography}
\end{document}